\documentclass{icsm}          
\usepackage{epsfig}
\begin{document}
\title{On a group-theoretical approach to the periodic table of chemical elements}
\authori{Maurice R. Kibler}      
\addressi{Institut de Physique Nucl\'eaire de Lyon \\
IN2P3-CNRS et Universit\'e Claude Bernard \\ 
43 Bd du 11 Novembre 1918, F-69622 Villeurbanne Cedex, France}
\authorii{}     \addressii{}
\authoriii{}    \addressiii{}
\authoriv{}     \addressiv{}
\authorv{}      \addressv{}
\authorvi{}     \addressvi{}
\headauthor{M.R. Kibler}            
\headtitle{On the periodic table \ldots}             
\lastevenhead{M.R. Kibler: On the periodic table \ldots}

\maketitle

\begin{abstract}
This paper is concerned with the application of the group SO($4,2$)$\otimes$SU(2) 
to the periodic table of chemical elements. It is shown how the Madelung rule of 
the atomic shell model can be used for setting up a periodic table that can be 
further rationalized via the group SO($4,2$)$\otimes$SU(2) and some of its 
subgroups. Qualitative results are obtained from the table and the 
general lines of a programme for a quantitative approach to the 
properties of chemical elements are developed on the basis of 
the group SO($4,2$)$\otimes$SU(2). 
\end{abstract}

\section{Introduction}

Most of the modern presentations of the periodic table of chemical elements are 
based on a quantum-mechanical treatment of the atom. In this respect, the simplest 
atom, namely the hydrogen atom, often constitutes a starting point for studying 
many-electron atoms. Naively, we may expect to construct an atom with atomic 
number $Z$ by distributing $Z$ electrons on the one-electron energy levels of a 
hydrogen-like atom. This building-up principle can be rationalized and refined from 
a group-theoretical point of view. As a matter of fact, we know that the dynamical 
noninvariance group of a hydrogen-like atom is the special real pseudo-orthogonal 
group in 4+2 dimensions (or conformal group) SO($4,2$) or SO($4,2$)$\otimes$SU(2) 
if we introduce the group SU(2) that labels the spin [1-3]. This result 
can be derived in several ways. We briefly review two of them. 

The first way is quite well-known and corresponds to a symmetry ascent process starting from 
the geometrical symmetry group SO(3) of a hydrogen-like atom. Then, we go from SO(3) 
to the dynamical invariance group SO(4) for the discrete spectrum or SO($3,1$) for 
the continuous spectrum. The relevant quantum numbers for the discrete spectrum are   
$n$, $\ell$ and $m_{\ell}$ (with $n=1, 2, 3, \cdots$; 
for fixed $n$:       $\ell = 0, 1, \cdots, n-1$; 
for fixed $\ell$: $m_{\ell} = -\ell, -\ell+1, \cdots, \ell$). The corresponding 
state vectors $\Psi_{n \ell m_{\ell}}$ can be organized to span multiplets of SO(3) 
and SO(4). The set 
\{$\Psi_{n \ell m_{\ell}}: n \ {\rm and} \ \ell \ {\rm fixed}; m_{\ell} \ {\rm ranging}$\}
generates an irreducible representation class (IRC), noted ($\ell$), while the set 
\{$\Psi_{n \ell m_{\ell}}: n \ {\rm fixed}; \ell \ {\rm and} \ m_{\ell} \ {\rm ranging}$\}
generates an IRC of SO(4). The direct sum 
$$
h = \bigoplus_{n=1}^{\infty} \bigoplus_{\ell=0}^{n-1} (\ell) 
$$
spanned by all the possible state vectors $\Psi_{n \ell m_{\ell}}$ corresponds to an
IRC of the de Sitter group SO($4,1$). The IRC $h$ is also an IRC of SO($4,2$). This IRC 
thus remains irreducible when restricting SO($4,2$) to SO($4,1$) but splits into two 
IRC's when restricting SO($4,2$) to SO($3,2$). The groups SO($4,2$), SO($4,1$) and SO($3,2$)
are dynamical noninvariance groups in the sense that not all their generators 
commute with the Hamiltonian of the hydrogen-like atom.    

The second way to derive SO($4,2$) corresponds to a symmetry descent process starting 
from the dynamical noninvariance group Sp($8,{\bf R}$), the real symplectic group in 
8 dimensions, for a four-dimensional isotropic 
harmonic oscillator. We know that there is a connection between the hydrogen-like atom 
in ${\bf R}^3$ and a four-dimensional oscillator in ${\bf R}^4$ [3-4]. Such a connection 
can be established via Lie-like methods (local or infinitesimal approach) or algebraic 
methods based on the so-called Kustaanheimo-Stiefel transformation (global or partial 
differential equation approach). Both approaches give rise to a constraint and the 
introduction of this constraint into the Lie algebra of Sp($8,{\bf R}$) produces a Lie 
algebra under constraints that turns out to be isomorphic with the Lie algebra of SO($4,2$). 
From a mathematical point of view, the latter Lie algebra is given by [5]
$$
{ \rm cent }_{ { \rm sp }(8,{\bf R}) } { \rm so(2) } / { \rm so(2) } = 
{\rm su(2,2)} \sim {\rm so(4,2)}
$$
in terms of Lie algebras.

Once we accept that the hydrogen-like atom may serve as a guide for studying the periodic 
table, the group SO($4,2$) and some of its sugroups play an important role in the construction
of this table. This was first realized in by Barut [6] and, independently, by 
Konopel'chenko [7]. Later, 
Byakov {\it et al}. [8]
further developed this group-theoretical approach of the periodic chart of chemical 
elements by introducing the direct product SO($4, 2$)$\otimes$SU(2). 

The aim of this paper is to emphasize the importance of a connection between the periodic table {\it \`a la} 
SO($4,2$)$\otimes$SU(2) and the Madelung rule [9] of atomic spectroscopy. The material is 
organized as follows. In Section 2, a periodic table is derived from the Madelung rule. 
In Section 3, this table is rationalized in terms of SO($4,2$)$\otimes$SU(2). Some 
qualitative and quantitative aspects of the table are described in Sections 4 and 5. This
paper constitutes a companion article to four previous papers by the author [10].   
           
\section{The periodic table {\it \`a la} Madelung}               

We first describe the construction of a periodic table based on the so-called Madelung 
rule [9] which arises from the atomic shell model. This approach to the periodic 
table uses the quantum numbers occurring in the quantum-mechanical treatment of the 
hydrogen atom as well as of a many-electron atom. 

In the central-field approximation, 
each of the $Z$ electrons of an atom with atomic number $Z$ is partly characterized by 
the quantum numbers $n$, $\ell$, and $m_{\ell}$. The numbers $\ell$ and $m_{\ell}$ 
are the orbital quantum 
number and the magnetic quantum number, respectively. They are connected to the chain 
of groups SO(3)$\supset$SO(2): the quantum number 
$\ell$ characterizes an IRC, of dimension $2 \ell + 1$, of SO(3) and $m_{\ell}$ a 
one-dimensional IRC of SO(2). The principal quantum number $n$ is such that $n - \ell - 1$ 
is the number of nodes of the radial wave function associated with the doublet $(n, \ell)$. In 
the case of the hydrogen atom or of a 
hydrogen-like atom, the number $n$ is connected to the group SO(4): the quantum number $n$ 
characterizes an IRC, of dimension $n^2$, of SO(4). The latter IRC of SO(4) splits into 
the IRC's of SO(3) corresponding to $\ell = 0, 1, \cdots, n-1$ when SO(4) is restricted to 
SO(3). A complete characterization of the dynamical state of each electron is provided by 
the quartet ($n, \ell, m_{\ell}, m_s$) or alternatively ($n, \ell, j, m$). Here, the spin 
$s = \frac{1}{2}$ of the electron has been introduced and $m_s$ is the $z$-component of 
the spin. Furthermore, $j = \frac{1}{2}$ for $\ell = 0$ and $j$ can take the values 
$j = \ell - s$ and $j = \ell + s$ for $\ell \not= 0$. The quantum numbers $j$ and $m$ are 
connected to the chain of groups SU(2)$\supset$U(1): $j$ characterizes an IRC, of 
dimension $2j+1$, of SU(2) and $m$ a one-dimensional IRC of U(1). 

Each doublet $(n, \ell)$ defines an atomic shell. The ground state of the atom is obtained 
by distributing the $Z$ electrons of the atom among the various atomic shells 
$n \ell, n' \ell', n'' \ell'', \cdots$ according to (i) an ordeing rule and (ii) the Pauli 
exclusion principle. A somewhat idealized situation is provided by the Madelung ordering 
rule: the energy of the shells increases with $n+\ell$ and, for a given value of $n+\ell$, 
with $n$ [9]. This may be depicted by Fig.~1 where the rows are labelled with 
$n = 1, 2, 3, \cdots$ and the columns with $\ell = 0, 1, 2, \cdots$ and where the entry in
the $n$-th row and $\ell$-th column is $[n + \ell, n]$. We thus have the ordering 
$[1, 1] < 
[2, 2] < 
[3, 2] < [3, 3] < 
[4, 3] < [4, 4] < 
[5, 3] < [5, 4] < [5, 5] < 
[6, 4] < [6, 5] < [6, 6] < 
\cdots$. This dictionary order corresponds to the following ordering of the $n \ell$ shells
$$
1s < 
2s < 
2p < 3s < 
3p < 4s < 
3d < 4p < 5s < 
4d < 5p < 6s < 
\cdots             
$$
which is verified to a good extent by experimental data. (According to atomic spectroscopy, 
we use the letters $s$ (for sharp), $p$ (for principal), $d$ (for diffuse), $f$ 
(for fundamental), $\dots$ to denote $\ell = 0, 1, 2, 3, \cdots$, respectively.)

\bfg
\bc
\epsfig{file=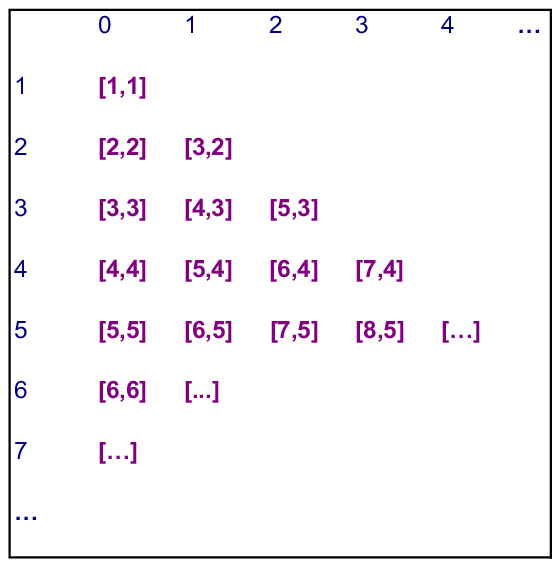}
\ec
\vspace{-5mm}
\caption{The $[n + \ell, n]$ Madelung array. The lines are labelled by 
$n = 1, 2, 3, \cdots$ and the columns by $\ell = 0, 1, 2, \cdots$. For 
fixed $n$, the label $\ell$ assumes the values $\ell = 0, 1, \cdots, n-1$.}
\efg

\bfg
\bc
\epsfig{file=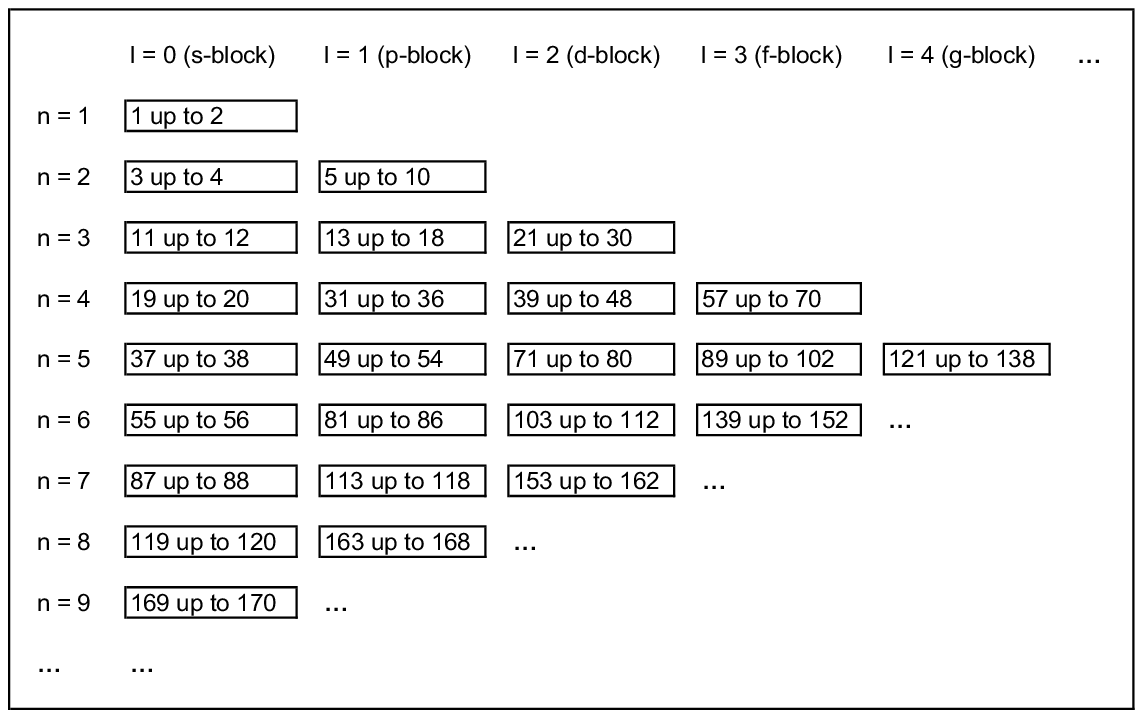}
\ec
\vspace{-5mm}
\caption{The periodic table deduced from the Madelung array. The box 
$[n + \ell, n]$ is filled with $2(2 \ell + 1)$ elements. The filling
of the various boxes $[n + \ell, n]$ is done according to the dictionary
order implied by Fig.~1.}
\efg

From these considerations of an entirely atomic character, we can construct a periodic 
table of chemical elements. We start from the Madelung array of Fig.~1. Here, the 
significance of the quantum numbers $n$ and $\ell$ is abandoned. The numbers $n$ and $\ell$ 
are now simple row and column indexes, respectively. We thus forget about the significance 
of the quartet $n, \ell, j, m$. The various blocks $[n + \ell, n]$ are filled in the 
dictionary order, starting from $[1, 1]$, with chemical elements of increasing atomic 
numbers. More precisely, the block $[n + \ell, n]$ is filled with $2(2 \ell + 1)$ elements, 
the atomic numbers of which increase from left to right. This yields Fig.~2, where each element 
is denoted by its atomic number $Z$. For instance, the block $[1, 1]$ is filled with 
$2(2 \times 0 + 1)=2$ elements corresponding to $Z=1$ up to $Z=2$. In a similar way, the blocks 
$[2, 2]$ and $[3, 2]$ are filled with $2(2 \times 0 + 1)=2$ elements and $2(2 \times 1 + 1)=6$ 
elements corresponding to $Z=3$ up to $Z=4$ and to $Z=5$ up to $Z=10$, respectively. It is to be 
noted, that the so obtained periodic table {\it a priori} contains an infinite number of 
elements: the $n$-th row contains $2n^2$ elements and each column (bounded from top) contains 
an infinite number of elements.

\section{The periodic table {\it \`a la} SO($4,2$)$\otimes$SU(2)}

We are now in a position to give a group-theoretical articulation to the periodic table of 
Fig.~2. For fixed $n$, the $2(2 \ell + 1)$ elements in the block $[n+\ell, n]$, that we shall 
refer to an $\ell$-block, may be labelled in the following way. For $\ell = 0$, the $s$-block 
in the $n$-th row contains two elements that we can distinguish by the number $m$ with $m$ 
ranging from $-\frac{1}{2}$ to $\frac{1}{2}$ when going from left to right in the row. For 
$\ell \not= 0$, the $\ell$-block in the $n$-th row can be divided into two sub-blocks, one 
corresponding to $j = \ell - \frac{1}{2}$ (on the left) and the other to 
$j = \ell + \frac{1}{2}$ (on the right). Each sub-block contains $2j+1$ elements 
(with $2j+1 = 2 \ell$       for $j = \ell - \frac{1}{2}$ 
 and  $2j+1 = 2 (\ell + 1)$ for $j = \ell + \frac{1}{2}$) that can be 
distinguished by the number $m$ with $m$ ranging from $-j$ to $j$ by step of one unit when going 
from left to right in the row. In other words, a chemical element can be located in the table 
by the quartet ($n, \ell, j, m$) (where $j = \frac{1}{2}$ for $\ell = 0$).

Following Byakov {\it et al}.~[8], it is perhaps interesting to use an image with streets, avenues and houses in 
a city. Let us call Mendeleev city the city whose (west-east) streets are labelled by $n$ and 
(north-south) avenues by ($\ell, j, m$). In the $n$-th street there are $n$ blocks of houses. The 
$n$ blocks are labelled by $\ell = 0, 1, \cdots, n-1$ so that the address of a block is ($n, \ell$). Each 
block contains one sub-block (for $\ell = 0$) or two sub-blocks (for $\ell \not= 0$). An address 
($n, \ell, j, m$) can be given to each 
house: $n$ indicates the street, $\ell$ the block, $j$ the sub-block and $m$ the location inside 
the sub-block. The organization of the city appears in Fig.~3. 

At this stage, it is worthwhile to re-give to the quartet ($n, \ell, j, m$) its group-theoretical 
significance. Then, Mendeleev city is clearly associated to the IRC class $h\otimes[2]$ of 
SO($4,2$)$\otimes$SU(2) where $[2]$ stands for the 
fundamental representation of SU(2). The whole city corresponds to the IRC
$$
       \bigoplus_{n=1}^{\infty} 
       \bigoplus_{\ell=0}^{n-1} 
       \bigoplus_{j=|\ell - \frac{1}{2}|}^{j=\ell + \frac{1}{2}} (j) 
=
\left( \bigoplus_{n=1}^{\infty} \bigoplus_{\ell=0}^{n-1} (\ell) \right) \otimes [2]
$$
of SO($4,2$)$\otimes$SU(2) in the sense that all the possible quartets ($n, \ell, j, m$), or alternatively 
($n, \ell, m_{\ell}, m_s$), can be associated to state vectors spanning this IRC. (In the latter 
equation, ($\ell$) and ($j$) stand for the IRC's of SO(3) and SU(2) associated 
with the quantum numbers $\ell$ and $j$, respectively.)   

We can ask the question: how to move in Mendeleev city? Indeed, there are several bus lines to go from 
one house to another one? The SO(3) bus lines (also called SO(3)$\otimes$SU(2) ladder operators) make it 
possible to go from one house in a given $\ell$-block to another house in the same $\ell$-block (see Fig.~4).
The SO(4)    bus lines (also called SO(4)$\otimes$SU(2) ladder operators) and 
the SO($2,1$) bus lines (also called SO($2,1$)            ladder operators) allow to move in a given street
(see Fig.~5) and in a given avenue (see Fig.~6), respectively. Finally, it should be noted that there are
taxis (also called SO($4,2$)$\otimes$SU(2) ladder operators) to go from a given house to an arbitrary house.
 
Another question concerns the inhabitants (also called chemical elements) of Mendeleev city. In fact, they 
are distinguished by a number $Z$ (also called the atomic number). The inhabitant living at the address 
($n, \ell, j, m$) has the number
\begin{eqnarray*}
Z(n \ell j m) &=& \frac{1}{6} (n + \ell) [(n + \ell)^2 - 1] + 
                 \frac{1}{2} (n + \ell + 1)^2 \\
	     & & - 
                 \frac{1}{4} [1 + (-1)^{n + \ell}] (n + \ell +1) - 
                 4 \ell (\ell + 1) + \ell + j(2 \ell + 1) + m - 1
\end{eqnarray*}
Each inhabitant may also have a nickname. All the inhabitants up to $Z=110$ have a nickname. For example, 
we have Ds, or darmstadtium in full, for $Z=110$. Not all the houses in Mendeleev city are inhabited. The 
inhabited houses go from $Z=1$ to $Z=116$ (the houses $Z=113$ and $Z=115$ are occupied 
since the begining of 2004 [11]). The houses corresponding to $Z \ge 117$ are not presently 
inhabited. (When a house is not inhabited, we also say that the corresponding element has not been observed 
yet.) The houses from $Z=111$ to $Z=116$ are inhabited but have not received a nickname yet. The various 
inhabitants known at the present time are indicated on Fig.~7.    

It is not forbidden to get married in Mendeleev city. Each inhabitant 
may get married with one or several inhabitants (including one or several 
clones). For example, we know H$_2$ (including H and its clone), 
HCl (including H and Cl), and H$_2$O (including O, H and its clone). 
However, there is a strict rule in the city: the assemblages or married inhabitants 
have to leave the city. They must live in another city and go to a 
city sometimes referred to as a molecular city. (Only the clones 
can stay in Mendeleev city.)

\section{Qualitative  aspects of the periodic table} 

Going back to Physics and Chemistry, we now describe Mendeleev city as 
a periodic table for chemical elements. We have obtained a table with 
rows and columns for which the $n$-th row contains $2n^2$ elements and 
the ($\ell, j, m$)-th column contains an infinite number of elements. A 
given column corresponds to a family of chemical analogs in the standard 
periodic table and a given row may contain several periods of the standard 
periodic table. 

The chemical elements in their ground state  
are considered as different states of atomic matter:  
each atom in the table appears as a particular 
partner for the (infinite-dimensional) unitary irreducible 
representation $h\otimes[2]$ of the group 
SO($4,2$)$\otimes$SU(2), where SO($4,2$) is reminiscent of 
the hydrogen atom and SU(2) is introduced for a doubling 
purpose. In fact, it is 
possible to connect two partners of the representation 
$h\otimes[2]$ by making use of shift operators of the Lie 
algebra of SO($4,2$)$\otimes$SU(2). In other words, it is 
possible to pass from one atom to another one by means of 
raising or lowering operators. The internal dynamics of each 
element is ignored. In other words, each neutral atom is 
assumed to be a noncomposite physical system. By way of 
illustration, we give a brief description 
of some particular columns and rows of the table. 

The alkali-metal atoms (see Fig.~8) are in the first column (with $\ell=0$, $j=\frac{1}{2}$, 
and $m=-\frac{1}{2}$); in the atomic shell model, they correspond to an external 
shell of type $1s$, $2s$, $3s$, $\cdots$; we note that hydrogen (H)
belongs to the alkali-metal atoms. The second column (with $\ell=0$, $j=\frac{1}{2}$, 
and $m=\frac{1}{2}$) concerns the alkaline earth metals (see Fig.~9) with an external 
atomic shell of type $1s^2$, $2s^2$, $3s^2$, $\cdots$; we note that helium (He) belongs 
to the alkaline earth metals. The sixth column corresponds to chalcogens (see Fig.~10) 
and the seventh column to halogens (see Fig.~11); it is to be observed that hydrogen does not 
belong to halogens as it is often the case in usual periodic tables. The eighth column 
(with $\ell=1$, $j=\frac{3}{2}$, and $m=\frac{3}{2}$) gives the noble gases (see Fig.~12) 
with an external atomic shell of type $2p^6$, $3p^6$, $4p^6$, $\cdots$; helium (with the 
atomic configuration $1s^22s^2$) does not belong to the noble gases in contrast with 
usual periodic tables. 

The $d$-blocks with $n=3$, 4 and 5 yield the three familiar transition series (see Fig.~13): 
the iron group goes from Sc(21) to Zn(30), the palladium group from Y(39) to Cd(48) and the
platinum group from Lu(71) to Hg(80). A fourth transition series goes from Lr(103) to $Z=112$ 
(observed but not named yet). In the shell model, the four transition series 
correspond to the filling of the $nd$ shell while the $(n+1)s$ shell is fully occupied, with
$n=3$ (iron group series), $n=4$ (palladium group series), $n=5$ (platinum group series) and 
$n=6$ (fourth series). The two familiar 
inner transition series are the $f$-blocks with $n=4$ and $n=5$ (see Fig.~14): the lanthanide series 
goes from La(57) to Yb(70) and the actinide series from Ac(89) to No(102). Observe that lanthanides 
start with La(57) not Ce(58) and actinides start with Ac(89) not Th(90). We note that lanthanides 
and actinides occupy a natural place in the table and are not reduced to appendages as it is 
generally the case in usual periodic tables in 18 columns. A superactinide series is 
predicted to go from $Z=139$ to $Z=152$ (and not from $Z=122$ to $Z=153$ as predicted by Seaborg 
[12]). In a shell model approach, the inner transition series 
correspond to the filling of the $nf$ shell while the $(n+2)s$ shell is fully occupied, with
$n=4$ (lanthanides), $n=5$ (actinides) and $n=6$ (superactinides). The table in Fig.~7 shows that 
the elements from $Z=121$ to $Z=138$ form a new period having no homologue among the known elements. 

In Section 3, we have noted that each $\ell$-block with $\ell \not=0$ gives rise to two sub-blocks. As an 
example, the $f$-block for the lanthanides is composed of a sub-block (corresponding to $j=\frac{5}{2}$) 
from La(57) to Sm(62) and another one (corresponding to $j=\frac{7}{2}$) from Eu(63) to Yb(70) (see Fig.~15). 
This division corresponds to the classification in light or ceric rare earths and heavy or yttric rare earths. 
It has received justifications both from the experimental and the theoretical sides.

\bfg[p]
\bc
\epsfig{file=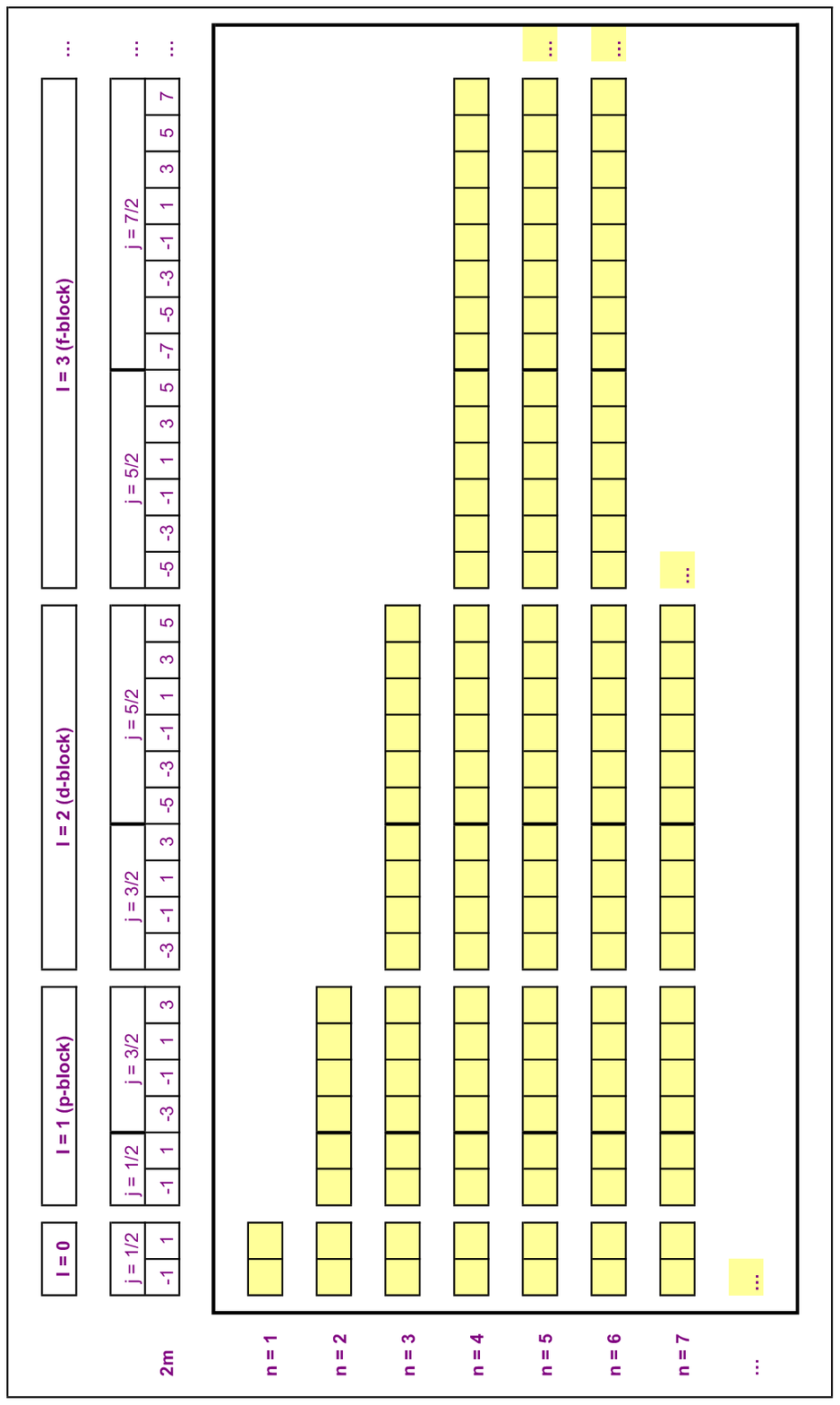}
\ec
\vspace{-5mm}
\caption{Mendeleev city. The streets are labelled by $n \in {\bf N}^*$ 
and the avenues by $(\ell, j, m)$ [$\ell = 0, 1, \cdots, n-1$;
$j=\frac{1}{2}$ for $\ell=0$, $j=\ell - \frac{1}{2}$ or $j=\ell + \frac{1}{2}$ 
for $\ell \not=0$; $m = -j, -j+1, \cdots, j$].}
\efg

\bfg[p]
\bc
\epsfig{file=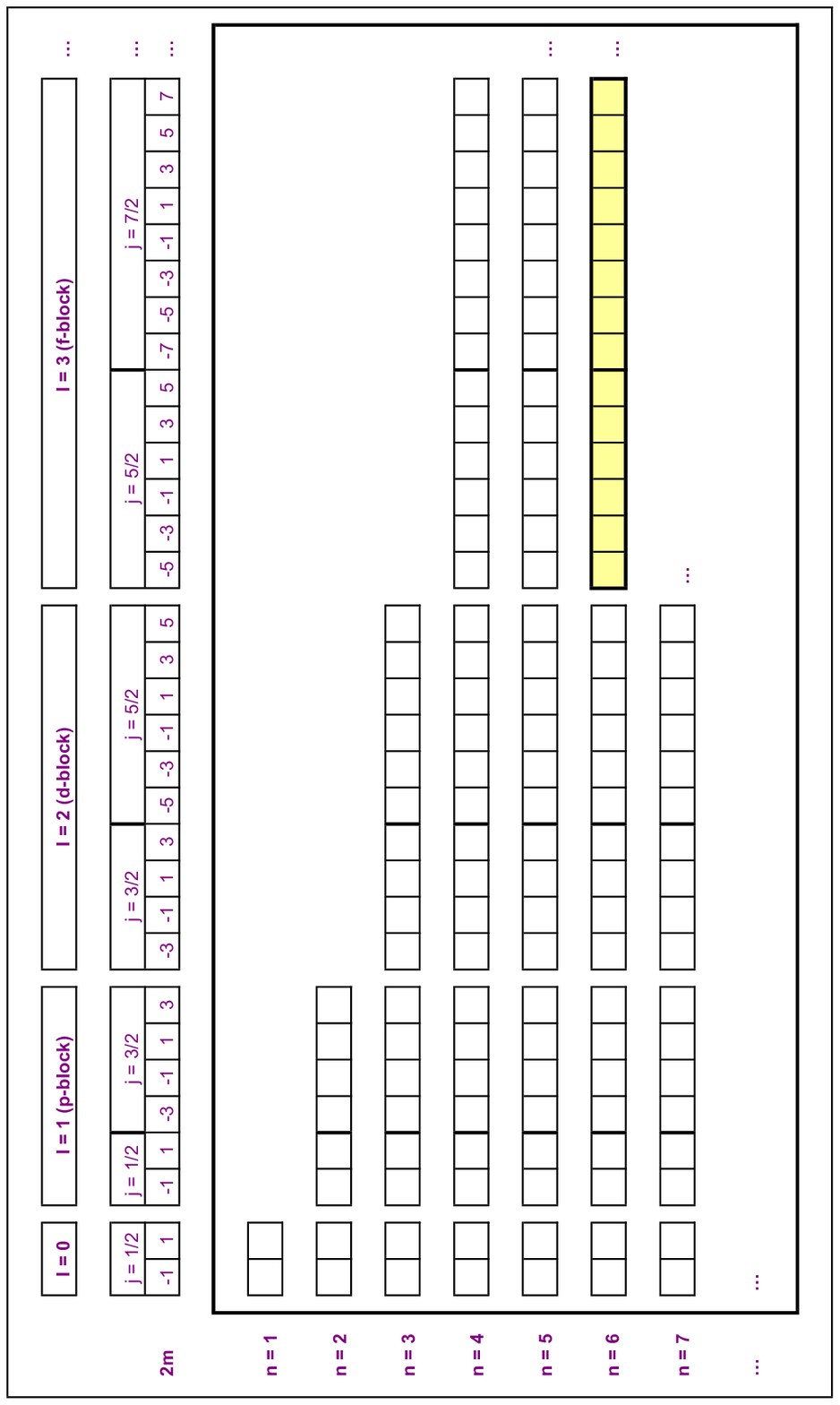}
\ec
\vspace{-5mm}
\caption{How to move in Mendeleev city? To move inside a block, take an
SO(3)$\otimes$SU(2) bus line! The shaded bus line corresponds to $\ell=3$ and 
$n=6$.}
\efg

\bfg[p]
\bc
\epsfig{file=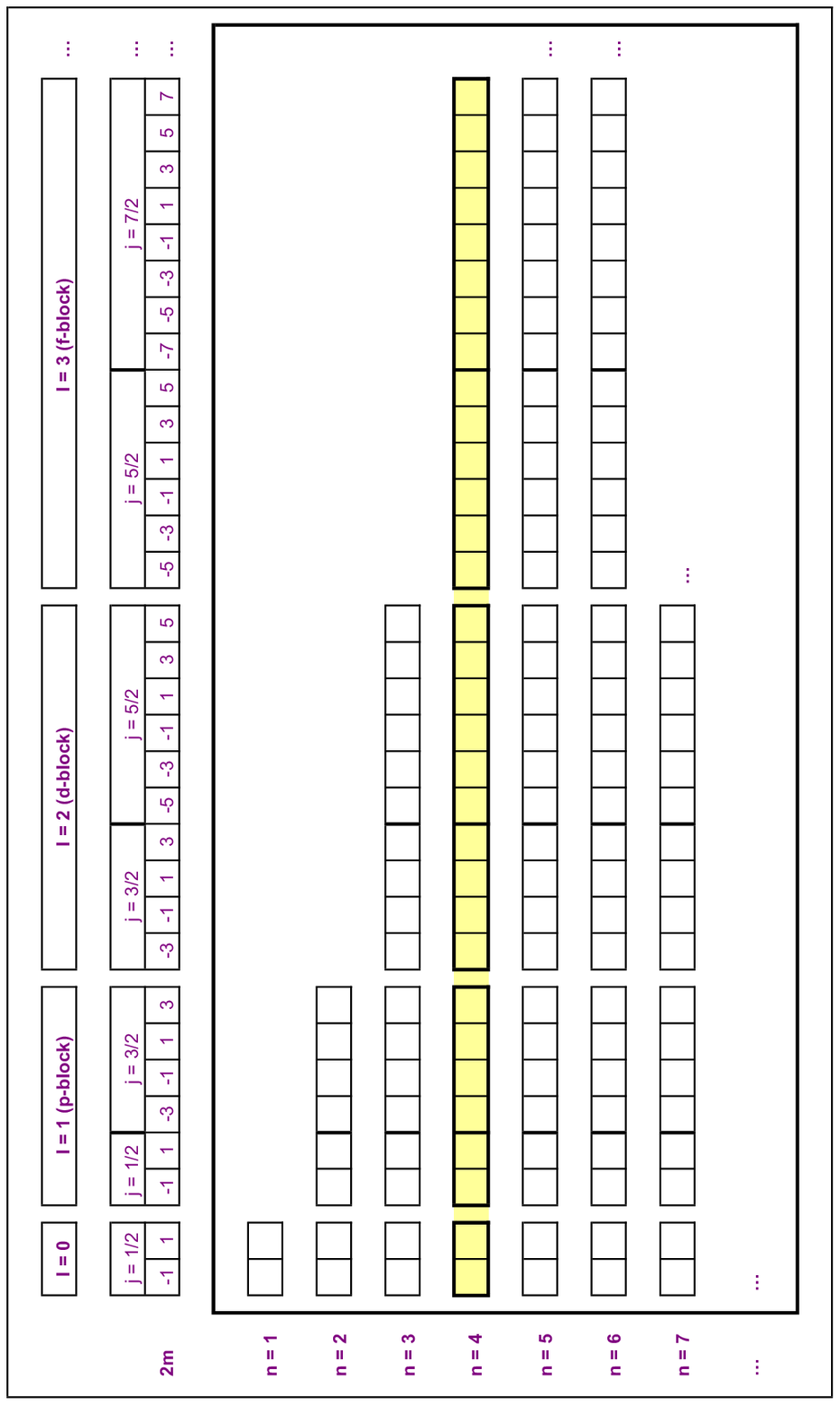}
\ec
\vspace{-5mm}
\caption{How to move in Mendeleev city? To move inside a west-east street, 
take an SO(4)$\otimes$SU(2) bus line! The shaded bus line corresponds to $n=4$.}
\efg

\bfg[p]
\bc
\epsfig{file=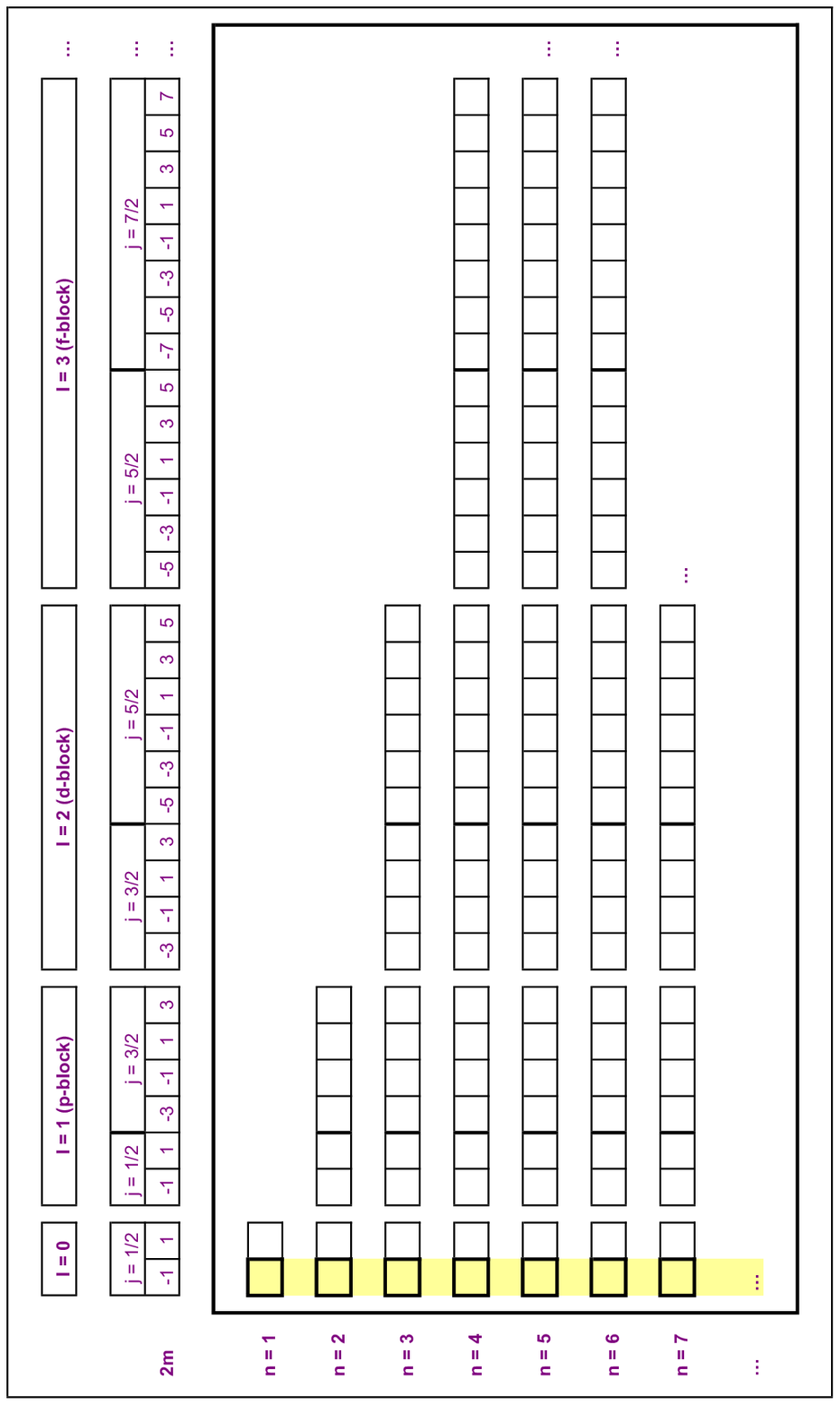}
\ec
\vspace{-5mm}
\caption{How to move in Mendeleev city? To move inside a north-south avenue, 
take an SO($2,1$) bus line! The shaded bus line corresponds to $\ell=0$, 
$j=\frac{1}{2}$, and $m=-\frac{1}{2}$.}
\efg

\bfg[p]
\bc
\epsfig{file=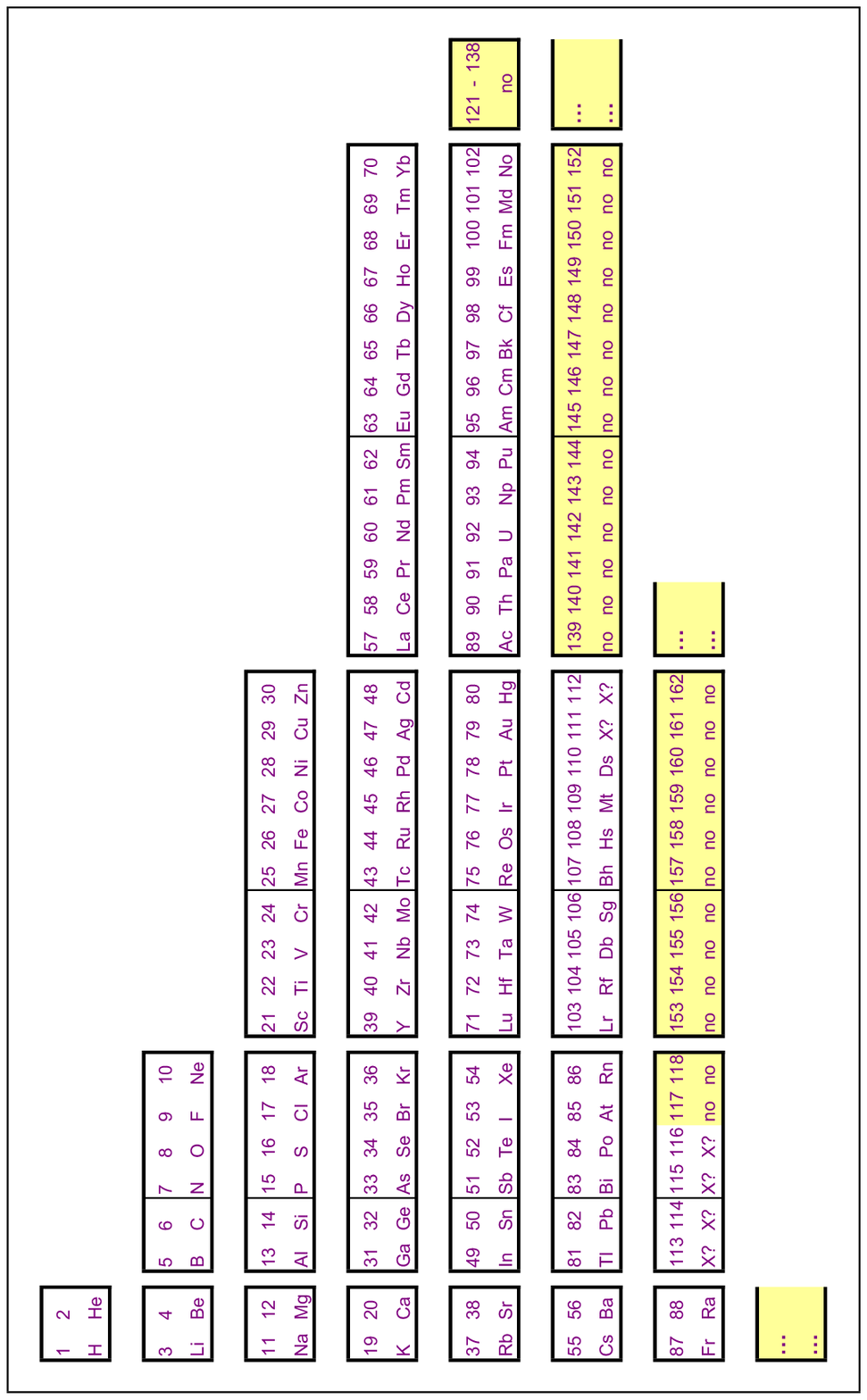}
\ec
\vspace{-5mm}
\caption{The inhabitants of Mendeleev city. The houses up to number $Z=116$ are 
inhabited [`X?' means inhabited (or observed) but not named, `no' means not
inhabited (or not observed)].}
\efg

\bfg[p]
\bc
\epsfig{file=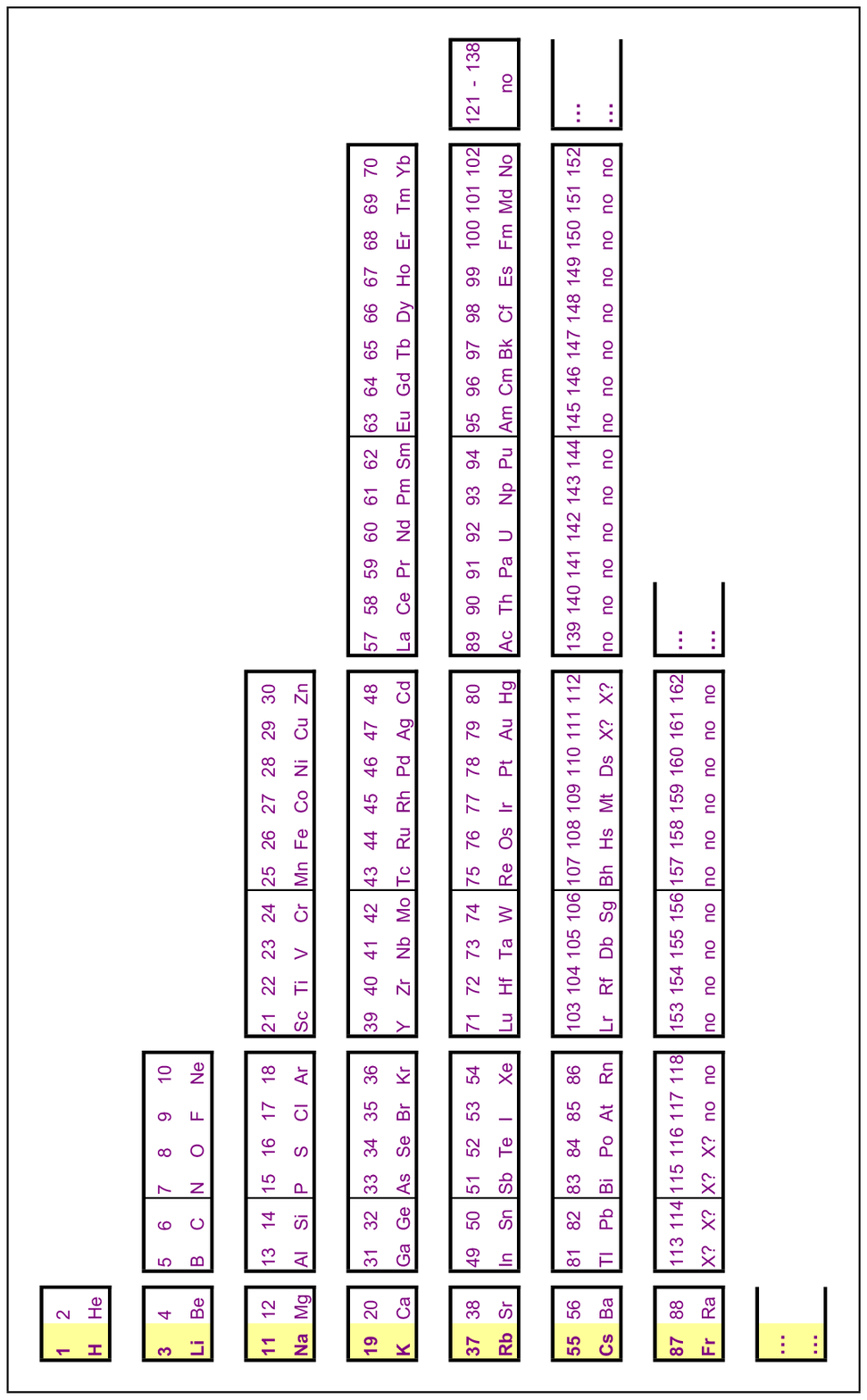}
\ec
\vspace{-5mm}
\caption{The inhabitants of Mendeleev city. The family of alkali metals 
(with an infinite number of elements) corresponds to 
($\ell=0$, $j=\frac{1}{2}$, $m=-\frac{1}{2}$) and $n \in {\bf N}^*$.}
\efg

\bfg[p]
\bc
\epsfig{file=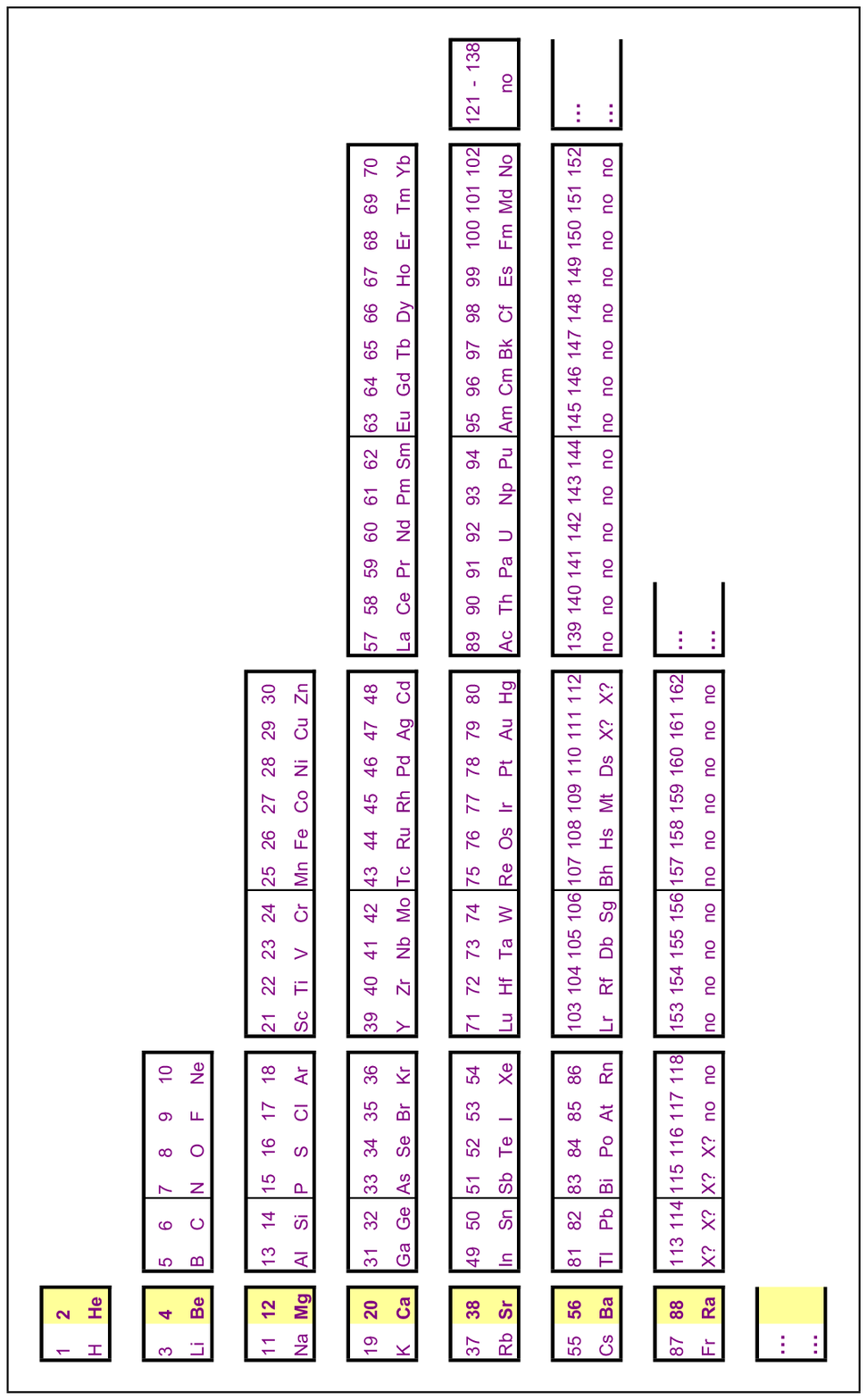}
\ec
\vspace{-5mm}
\caption{The inhabitants of Mendeleev city. The family of alkaline earth metals 
(with an infinite number of elements) corresponds to 
($\ell=0$, $j=\frac{1}{2}$, $m= \frac{1}{2}$) and $n \in {\bf N}^*$.}
\efg

\bfg[p]
\bc
\epsfig{file=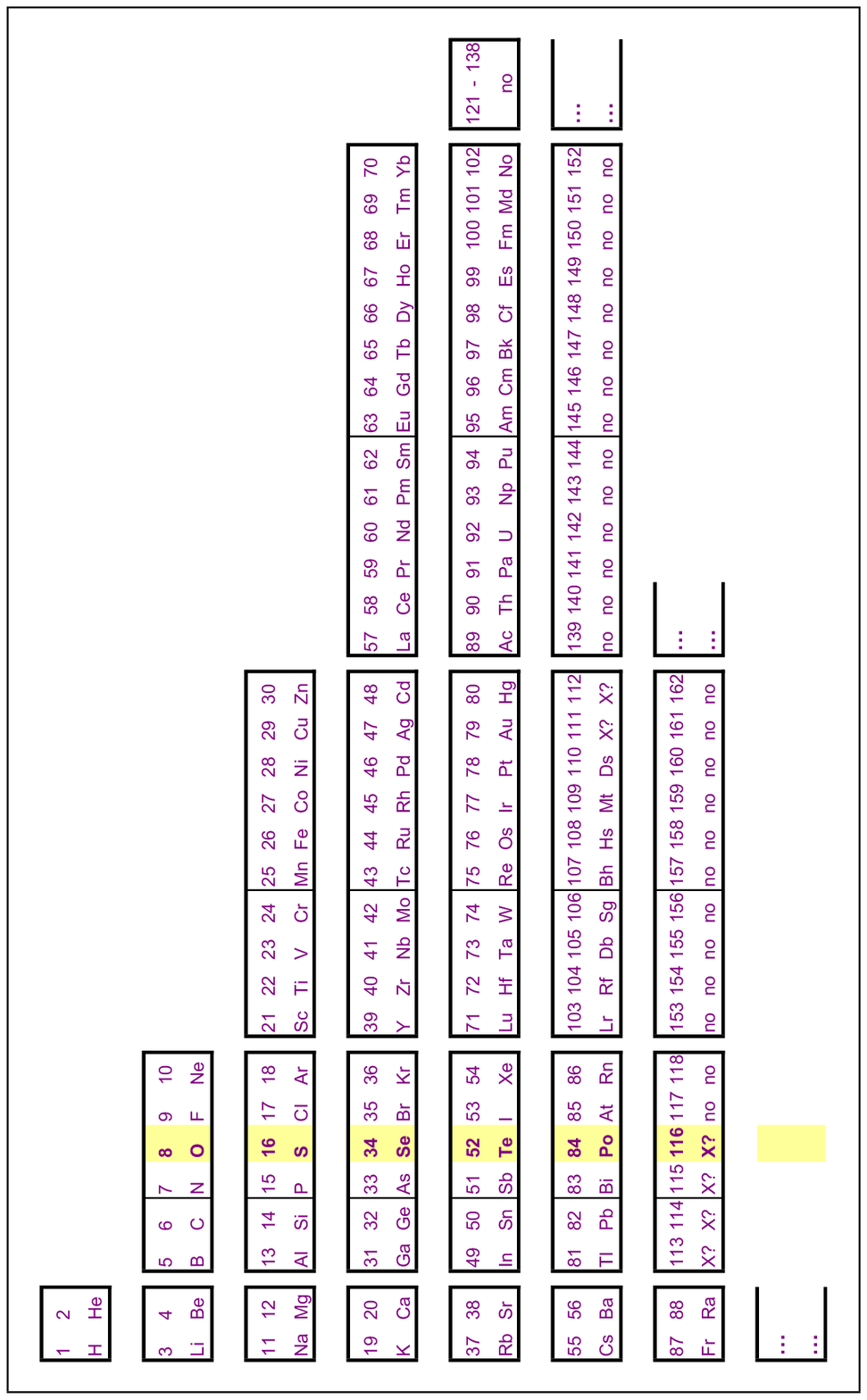}
\ec
\vspace{-5mm}
\caption{The inhabitants of Mendeleev city. The family of chalcogens 
(with an infinite number of elements) corresponds to 
($\ell=1$, $j=\frac{3}{2}$, $m=-\frac{1}{2}$) and $n = 2, 3, 4, \cdots$.}
\efg

\bfg[p]
\bc
\epsfig{file=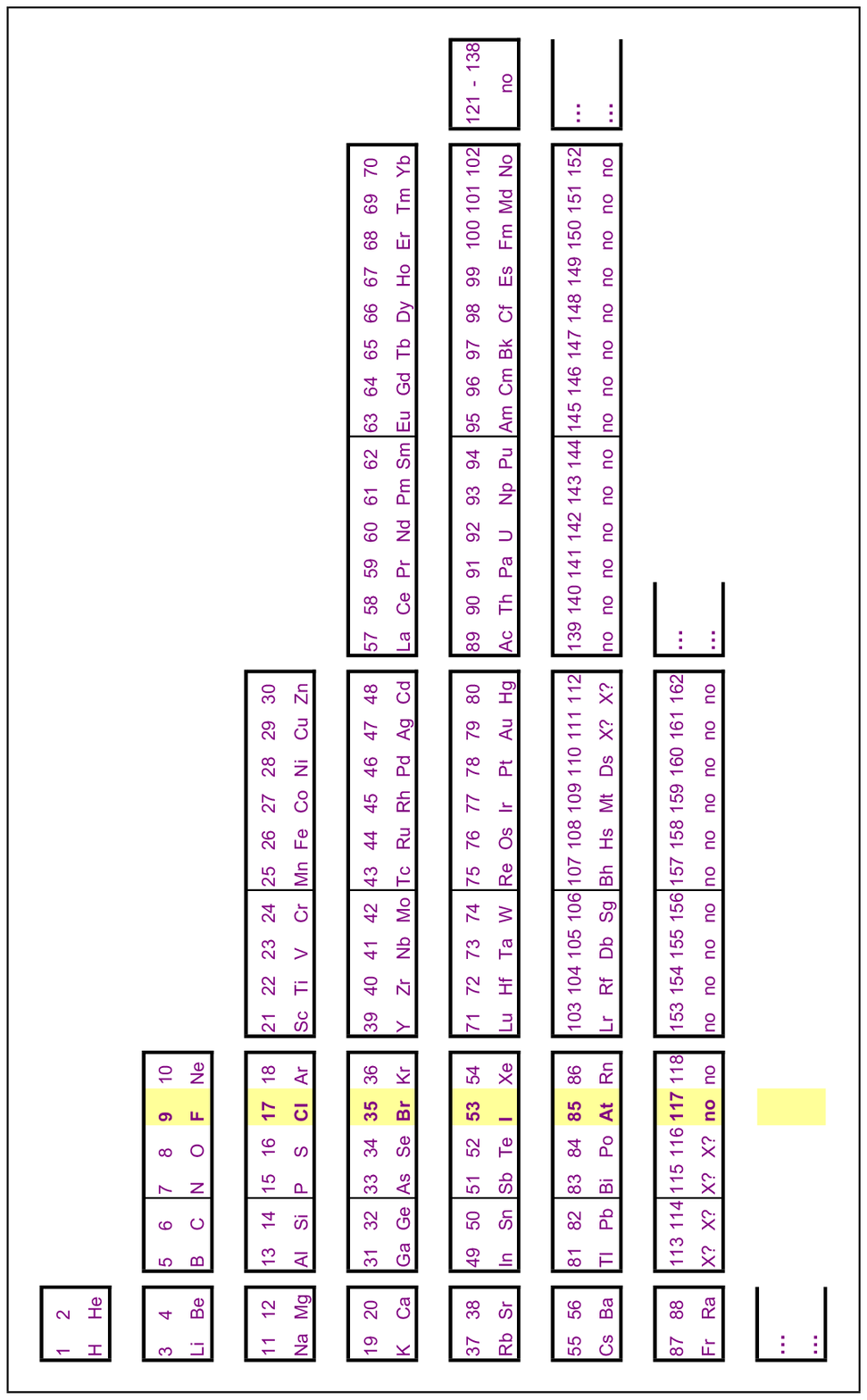}
\ec
\vspace{-5mm}
\caption{The inhabitants of Mendeleev city. The family of halogens 
(with an infinite number of elements) corresponds to 
($\ell=1$, $j=\frac{3}{2}$, $m= \frac{1}{2}$) and $n = 2, 3, 4, \cdots$.}
\efg

\bfg[p]
\bc
\epsfig{file=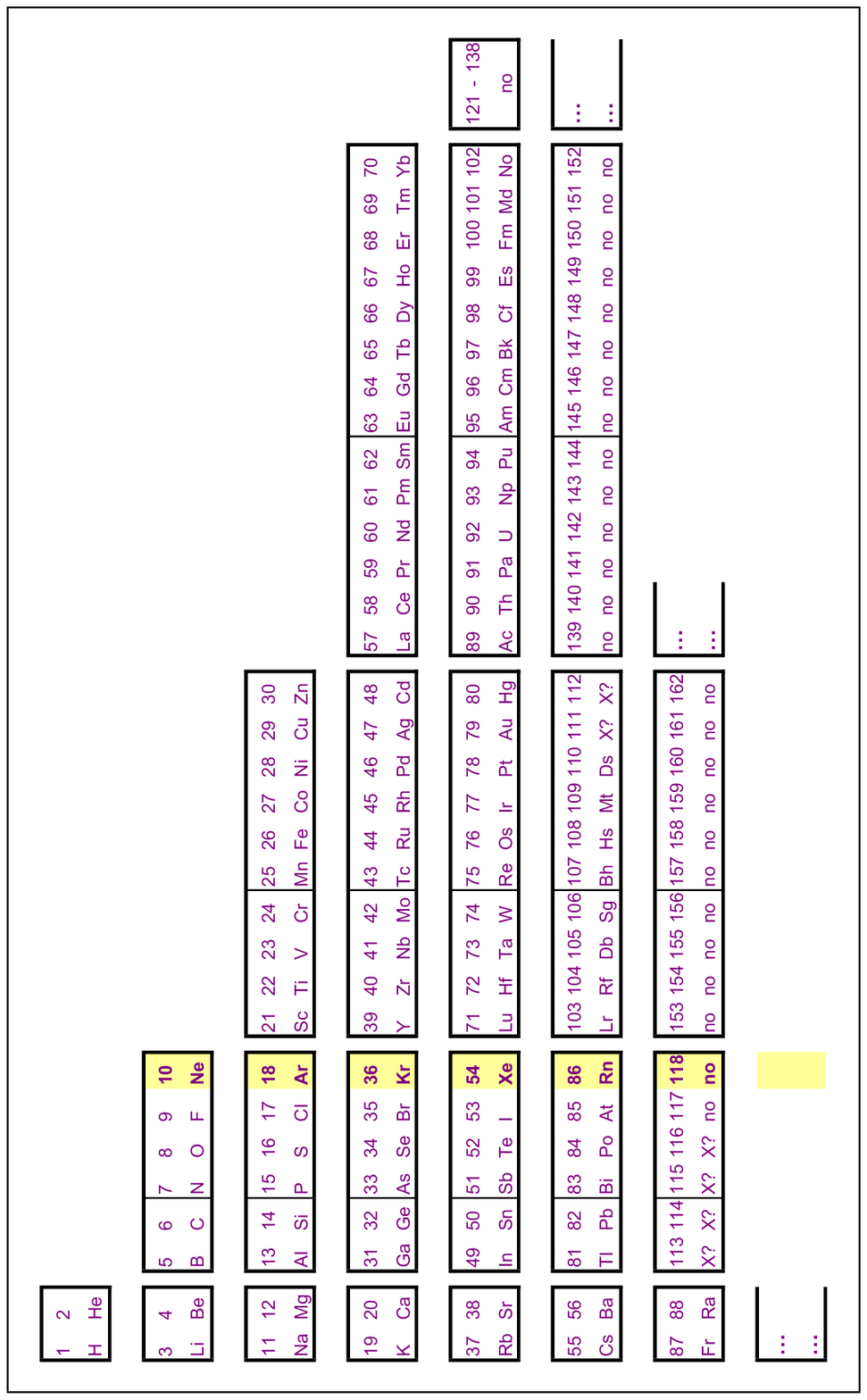}
\ec
\vspace{-5mm}
\caption{The inhabitants of Mendeleev city. The family of noble gases 
(with an infinite number of elements) corresponds to 
($\ell=1$, $j=\frac{3}{2}$, $m= \frac{3}{2}$) and $n = 2, 3, 4, \cdots$.}
\efg

\bfg[p]
\bc
\epsfig{file=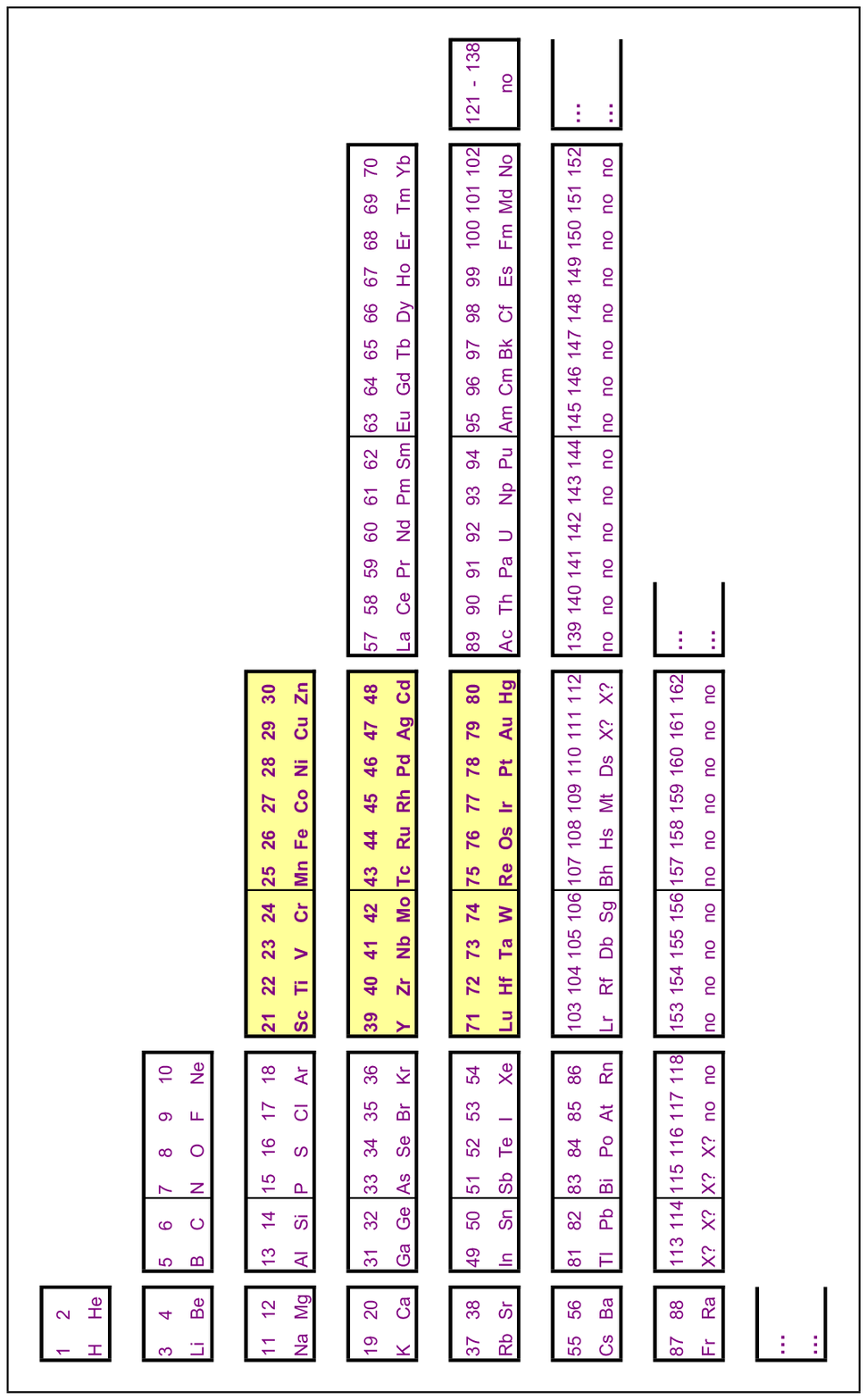}
\ec
\vspace{-5mm}
\caption{The inhabitants of Mendeleev city. The three transition series: 
the iron      group series [from Sc(21) to Zn(30)],
the palladium group series [from Y(39)  to Cd(48)], and
the platinum  group series [from Lu(71) to Hg(80)].}
\efg

\bfg[p]
\bc
\epsfig{file=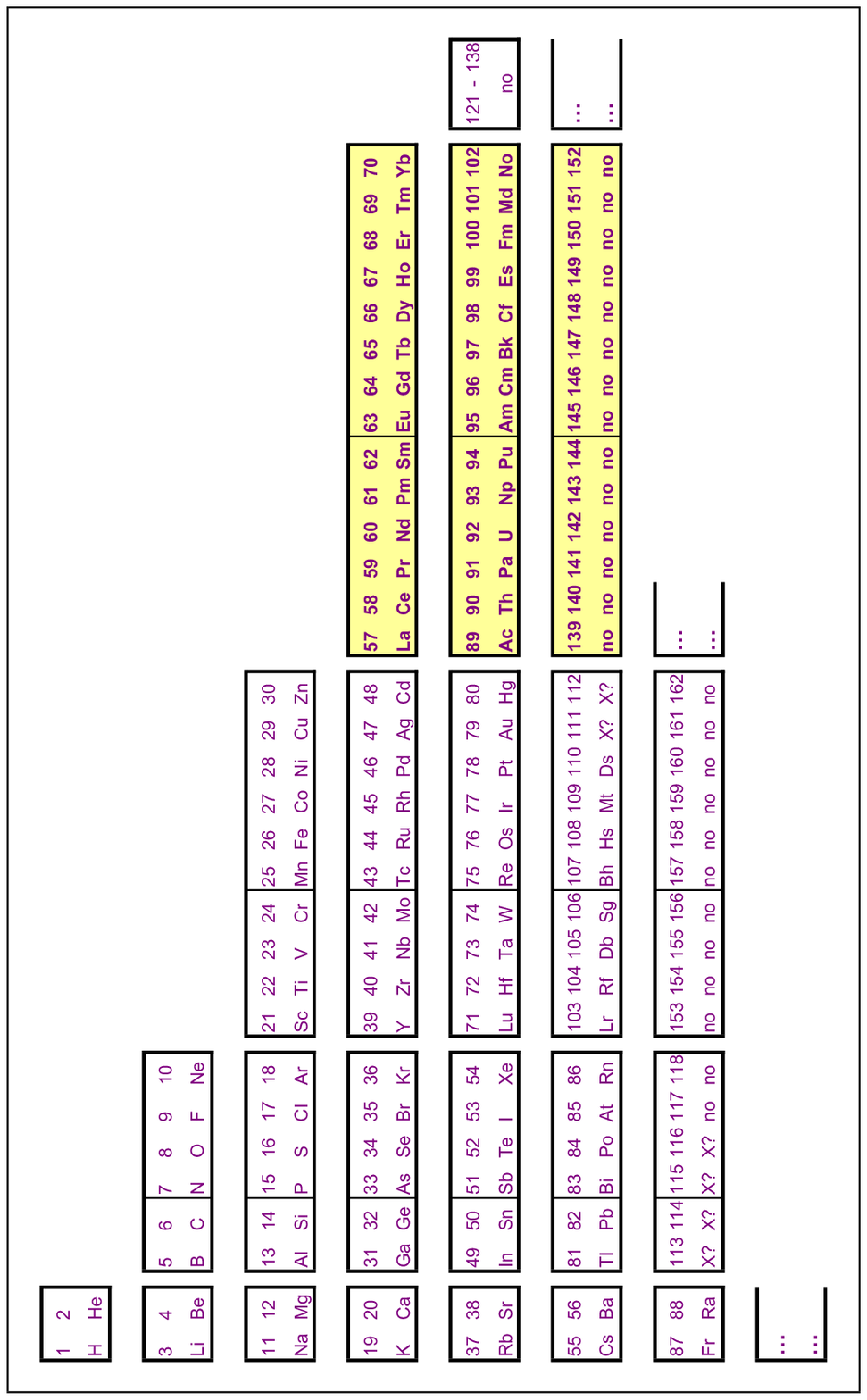}
\ec
\vspace{-5mm}
\caption{The inhabitants of Mendeleev city. The three inner transition series: 
the lanthanide      series [from La(57) to Yb(70)],
the actnide         series [from Ac(89) to No(102)], and
the superactinides series [from (139)  to (152)].}
\efg

\bfg[p]
\bc
\epsfig{file=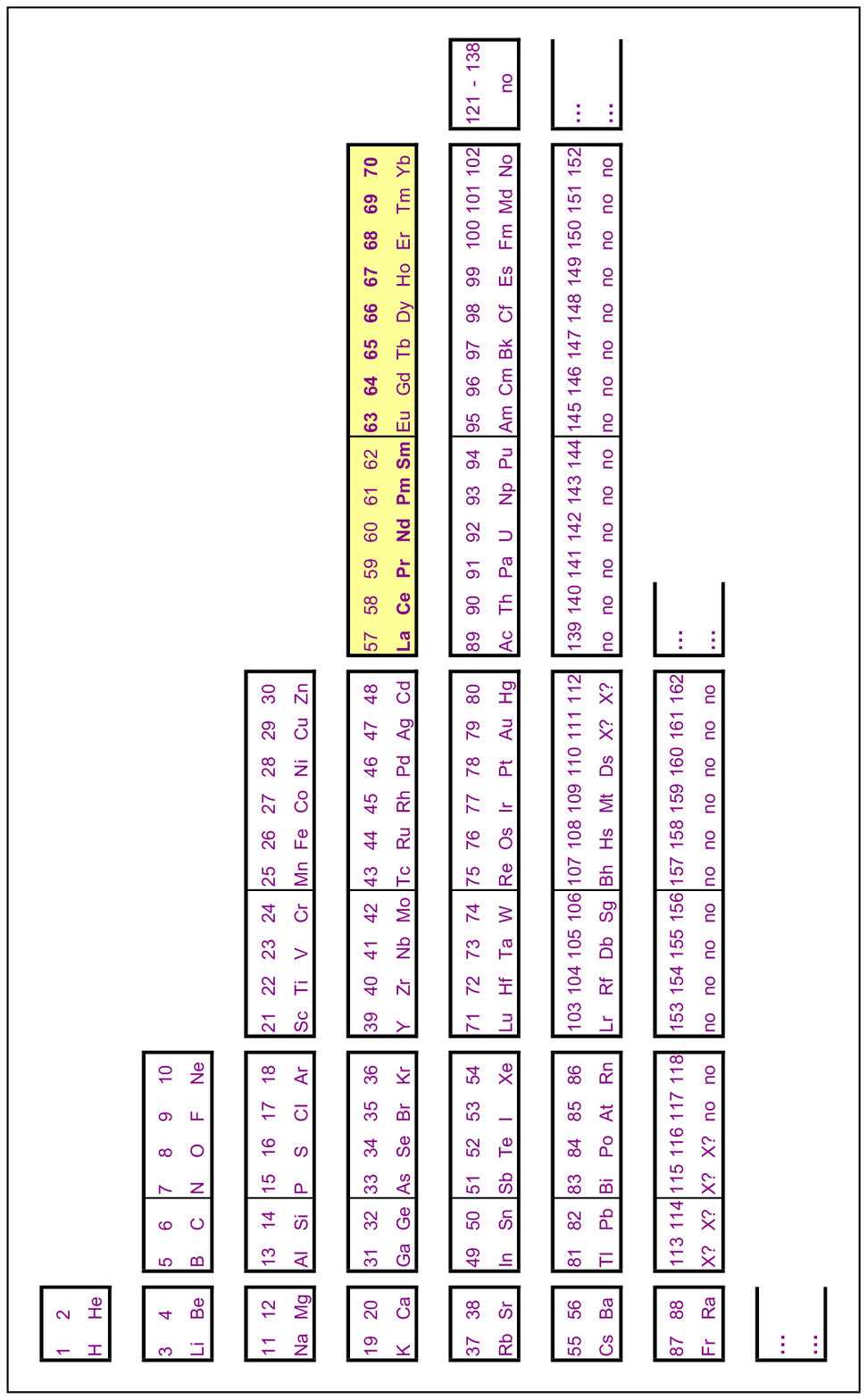}
\ec
\vspace{-5mm}
\caption{The inhabitants of Mendeleev city. The lanthanides or rare earths 
can be divided into two sub-blocks: 
the light (or ceric)  rare earths [from La(57) to Sm(62)] correspond 
to $j=\frac{5}{2}$ and 
the heavy (or yttric) rare earths [from Eu(63) to Yb(70)] 
to $j=\frac{7}{2}$.}
\efg

\section{Quantitative aspects of the periodic table}

To date, the use of SO($4,2$) or SO($4,2$)$\otimes$SU(2) in 
connection with periodic charts has been limited to 
qualitative aspects only, viz., classification of neutral 
atoms and ions as well. We would like to give here the main 
lines of a programme under development (inherited from nuclear physics and 
particle physics) for dealing with quantitative aspects.

The first step concerns the mathematics of the programme. The 
direct product group SO($4,2$)$\otimes$SU(2) 
is a Lie group of order eighteen. Let 
us first consider the SO($4,2$) part which is a semi-simple 
Lie group of order $r = 15$ and of rank $\ell = 3$. It has thus 
fifteen generators involving three Cartan generators (i.e., generators 
commuting between themselves). Furthermore, it has three invariant 
operators or Casimir operators (i.e., independent polynomials, 
in the enveloping algebra of the Lie algebra of SO($4,2$), that 
commute with all generators of the group SO($4,2$)). Therefore, 
we have a set of six ($3 + 3$) operators that commute between 
themselves: the three Cartan generators and the three Casimir operators. 
Indeed, this set is not complete from the mathematical point of 
view. In other words, the eigenvalues of the six above-mentioned 
operators are not sufficient for labelling the state vectors in 
the representation space of SO($4,2$). According to a (not very 
well-known) result popularised by Racah, we need to find
$\frac{1}{2}(r - 3 \ell) = 3$
additional operators in order to complete the set of the six preceding 
operators. This yields a complete set of nine ($6 + 3$) commuting operators 
and this solves the state labelling problem for the group SO($4,2$). 
The consideration of the group SU(2) is trivial: SU(2) is a semi-simple Lie 
group of order $r = 3$ and of rank $\ell = 1$ so that $\frac{1}{2}(r - 3 \ell) = 0$ 
in that case. As a result, we end up with a complete set of eleven ($9 + 2$) 
commuting operators. It is to be stressed that this result constitutes the key 
and original starting point of the programme. 

The second step establishes contact with chemical physics. Each of the eleven  
operators can be taken to be self-adjoint and thus, from the quantum-mechanical 
point of view, can describe an observable. Indeed, four of the eleven operators, 
namely, the three Casimir operators of SO($4,2$) and the Casimir operator of SU(2),
serve for labelling the representation $h\otimes[2]$ of SO($4,2$)$\otimes$SU(2) 
for which the various chemical elements are partners. The seven remaining operators 
can thus be used for describing chemical and physical properties of the elements
like: 
ionization energy; 
oxydation degree; 
electron affinity; 
electronegativity; 
melting and boiling points; 
specific heat; 
atomic radius; 
atomic volume; 
density; 
magnetic susceptibility; 
solubility; etc. In most cases, this can be done by 
expressing a chemical observable 
associated with a given property in terms of the seven operators which serve as 
an integrity basis for the various observables. Each observable can be developed 
as a linear combination of operators constructed from the integrity basis. This is 
reminiscent of group-theoretical techniques used in nuclear and atomic 
spectroscopy (cf. the Interacting Boson Model) or in hadronic spectroscopy (cf. the
Gell-Mann/Okubo mass formulas for baryons and mesons). 

The last step is to proceed to a diagonalisation process and then to 
fit the various linear combinations to experimental data. This 
can be achieved through fitting procedures concerning either a period of elements 
(taken along a same line of the periodic table) or a 
family of elements (taken along a same column of the 
periodic table). For each property this will lead to 
a formula or phenomenological law that can 
be used in turn for making predictions concerning the chemical elements for 
which no data are available. In addition, it is hoped that this will shed 
light on regularities and well-known (as well as recently discovered) patterns 
of the periodic table. This programme, referred to as the KGR programme, was 
briefly presented at the 2003 Harry Wiener International Conference. It is 
presently under progress. 

\section{Closing remarks}

We close this paper with two remarks. Possible extensions of this work concern 
isotopes and molecules. The consideration of isotopes needs the introduction 
of the number of nucleons in the atomic nucleus. With such an introduction we 
have to consider other dimensions for Mendeleev city: the city is not anymore
restricted to spread in Flatland. Group-theoretical analyses 
of periodic systems of molecules can be achieved by considering direct 
products involving several copies of SO($4,2$)$\otimes$SU(2). Several 
works have been already devoted to this subject [13]. 

\bigskip
{Acknowledgments. The author is indebted to M.S. Antony for comments concerning the elements
$Z = 110$, $113$ and $115$ and to Yu.Ts. Oganessian for providing him with Ref.~[11]. He would like 
also to thank I. Berkes, M. Cox, R.A. Hefferlin and M. Laing for interesting discussions and 
correspondence.}

\end {document}